\documentclass[aps,footinbib,prb,reprint]{revtex4-1}
\usepackage{graphicx}
\usepackage{amssymb}
\usepackage{float}
\usepackage{hyperref}
\usepackage{amsmath}
\usepackage{comment}
\usepackage{slashed}
\usepackage{lipsum}
\usepackage[dvipsnames]{xcolor}
\usepackage{xcolor}   
\usepackage{xspace} 
\usepackage{bm}
\usepackage{xfrac}
\usepackage{lineno}
\usepackage[normalem]{ulem}
\usepackage[toc,page]{appendix}
\usepackage[capitalize]{cleveref}
\usepackage{nicefrac}
\usepackage{cases}

\newcommand*{\citen}{}
\DeclareRobustCommand*{\citen}[1]{%
  \begingroup
    \romannumeral-`\x 
    \setcitestyle{numbers}%
    \cite{#1}%
  \endgroup
}

\usepackage[T1]{fontenc}      
\usepackage{ae}

\usepackage{grffile}

\newcommand{\im}{\ensuremath{\mathrm{i}}}

\RequirePackage{fix-cm} 
\DeclareMathSizes{9.5}{9.5}{7.0}{5}

\begin{document}



\title{Momentum-Dependent Mass and AC Hall Conductivity of Quantum Anomalous Hall Insulators and Their Relation to the Parity Anomaly}

\author{Christian~Tutschku}
\affiliation{Institute for Theoretical Physics  and W\"urzburg-Dresden Cluster of Excellence ct.qmat,
Julius-Maximilians-Universit\"at W\"urzburg, 97074 W\"urzburg, Germany}

\author{Jan~B\"ottcher}
\affiliation{Institute for Theoretical Physics  and W\"urzburg-Dresden Cluster of Excellence ct.qmat,
Julius-Maximilians-Universit\"at W\"urzburg, 97074 W\"urzburg, Germany}
\author{Ren\'e~Meyer}
\affiliation{Institute for Theoretical Physics  and W\"urzburg-Dresden Cluster of Excellence ct.qmat,
Julius-Maximilians-Universit\"at W\"urzburg, 97074 W\"urzburg, Germany}

\author{E.~M.~Hankiewicz} 
\email{Corresponding author: ewelina.hankiewicz@physik.uni-wuerzburg.de}
\affiliation{Institute for Theoretical Physics  and W\"urzburg-Dresden Cluster of Excellence ct.qmat,
Julius-Maximilians-Universit\"at W\"urzburg, 97074 W\"urzburg, Germany}

\begin{abstract}
The Dirac mass of a two-dimensional QAH insulator is directly related to the parity anomaly of planar quantum electrodynamics, as shown initially in Phys.~Rev.~Lett.~51,~2077~(1983). In this work, we connect the additional momentum-dependent Newtonian mass term of a QAH insulator to the parity anomaly.
We reveal that in the calculation of the effective action, before renormalization, the Newtonian mass acts similar to a parity-breaking element of a high-energy regularization scheme. This calculation allows us to derive the finite frequency correction to the DC Hall conductivity of a QAH insulator.  We predict that the leading order AC correction contains a term proportional to the Chern number. This term originates from the Newtonian mass and can be measured via electrical or magneto-optical experiments. Moreover, we prove that the Newtonian mass significantly changes the resonance structure of the AC Hall conductivity in comparison to pure Dirac systems like graphene.
\end{abstract}

\maketitle


\section{Introduction}

\vspace{-.2cm}

The discovery of Dirac materials, such as topological insulators \cite{Kane05A,Bernevig06,kng07,fu07,hughes08,roy09,hasan08,Xia09,Chen178,Sullivan11} and Weyl or Dirac \cite{Wan11,xu11,Borisenko14,liu14a,liu14b,Neupane14} semimetals yields the possibility to measure  quantum anomalies in condensed matter systems. In particular, it was shown that two-dimensional Quantum Anomalous Hall (QAH) insulators, such as (Hg,Mn)Te quantum wells\cite{Liu08,Qi08} or magnetically 
doped (Bi,Sb)Te thin films \cite{Yu61,Chang13} are directly related to the parity anomaly of planar quantum electrodynamics (QED$_{2+1}$) \cite{Tutschku19}.
In a nutshell, this anomaly implies that in odd space-time dimensions
it is not possible to quantize an odd number of Dirac fermions at the same time in a gauge- and in a  parity symmetric manner \cite{Niemi83,Redlich83,Haldane88}.


A two-dimensional QAH insulator consists of two Chern insulators, a topologically trivial, as well as topologically non-trivial one. Similar to the Haldane model, each of those Chern insulators realizes the parity anomaly if its mass gap is closed at the topological phase transition \cite{Haldane88}. In general, each Chern insulator is characterized by Dirac-like physics in 2+1 dimensions and, in our system, has two different parity-breaking mass terms: A momentum independent (Dirac) mass, as well as a momentum dependent (Newtonian) mass\cite{Liu16}.  As their contribution to the Chern number does not vanish in the parity symmetric (zero mass) limit, both of these mass terms are directly related to the parity anomaly \cite{Lu10,Qi08}.

While the contribution of the Dirac mass to the effective action of a QED$_{2+1}$ systems was initially shown  in Refs.~[\citen{Niemi83,Redlich83}], 
the effective action of Chern insulator in the presence of a  Newtonian mass has not yet been analyzed. To bridge this gap, we explicitly derive in this work  the polarization operator of a Chern insulator. In the corresponding calculation for a pure, massless QED$_{2+1}$ system, the parity anomaly arises from the particular regularization of the infinite Dirac sea. Even for a massive Dirac system,
each gauge-invariant regularization 
scheme necessarily needs to break the parity-symmetry due to the underlying parity anomaly. Consequently, such a regularization  provides another $1/2$ to the otherwise half-quantized Chern number of a single Dirac fermion. For the Chern insulators introduced above, this requirement is already fulfilled due to the presence of the Newtonian mass term. 
From this perspective the Newtonian mass term, provided by the material, acts similar to a parity-breaking element of a mathematical regularization scheme \cite{Note13}
\footnotetext[13]{With 'similar' we mean that since the Newtonian mass is a real parameter of the theory, it does not vanish during renormalization. This latter requirement needs to be fulfilled by any mathematical regulator.}.
Inspired by this property, we compare the role of the Newtonian mass term for the calculation of the effective action of a Chern insulator to common parity-breaking  regularization schemes of QED$_{2+1}$. In particular, we show that for the calculation of the Hall conductivity the Newtonian mass can be seen as a  continuum version of a Wilson mass term \cite{Coste89}, whereas it acts significantly different than a Pauli-Villars (PV) mass \cite{Redlich84}, or the integration contour in a $\zeta$-function regularization \cite{schaposnik96}. With continuum we mean that the Newtonian mass does not vanish during renormalization when the artificial lattice spacing in the Wilson mass term vanishes.
Even though the Newtonian mass implies an integer quantized DC Hall conductivity, it does not remove the UV divergence in the effective action, as it is done by common higher derivative regularization schemes \cite{Nagahama1986,Nagahama1986b}.
Consequently, the effective action of a Chern insulator still requires regularization. Due to the fact that a QAH insulator consists of two Chern insulators, a topologically trivial as well as a topologically non-trivial one, it is possible to regularize the entire system in a gauge and parity symmetric manner.
We choose two Pauli-Villars fields, one for each Chern insulator, which are  related by time-reversal symmetry. Together, these fields compensate the parity anomaly associated to each single Chern insulator.

\vspace{-.1cm}

Moreover, we derive the non-quantized finite frequency corrections to the DC Hall conductivity during our calculation of the effective action. We show that the leading order AC correction contains a term which is proportional to the Chern number. This term originates from the Newtonian mass and can be measured by the leading order AC corrections to the DC Kerr and Faraday angles. In addition, we show that the Newtonian mass fundamentally changes the resonance structure of the AC Hall conductivity in comparison to pure Dirac systems,  as it can alter the mass gap in QAH insulators.\\

This work is organized as follows: In Sec.~II, we introduce two-dimensional QAH insulators.
In Sec.~III, we review how to quantize a classical field theory and how the parity anomaly arises in a gapless QED$_{2+1}$ system. 
In Sec.~IV, we derive the effective action of a two-dimensional QAH insulator. We separately calculate the polarization operators of each of the two Chern insulators, together defining the entire QAH system. In particular, we show the integer quantized DC Hall conductivity of a QAH insulator and  derive its non-quantized AC correction. In Sec.~V, we compare the role of the Newtonian mass term for the calculation of the effective action to common QED$_\mathrm{2+1}$ regularization schemes, which also ensure an integer quantized Chern number.	In Sec.~VI, we summarize our  results and give an outlook. 

\vspace{-.3cm}

\section{Model} \label{model}

\vspace{-.3cm}

\begin{figure}[t]
\vspace{.2cm}
\centering
\includegraphics[scale=0.2]{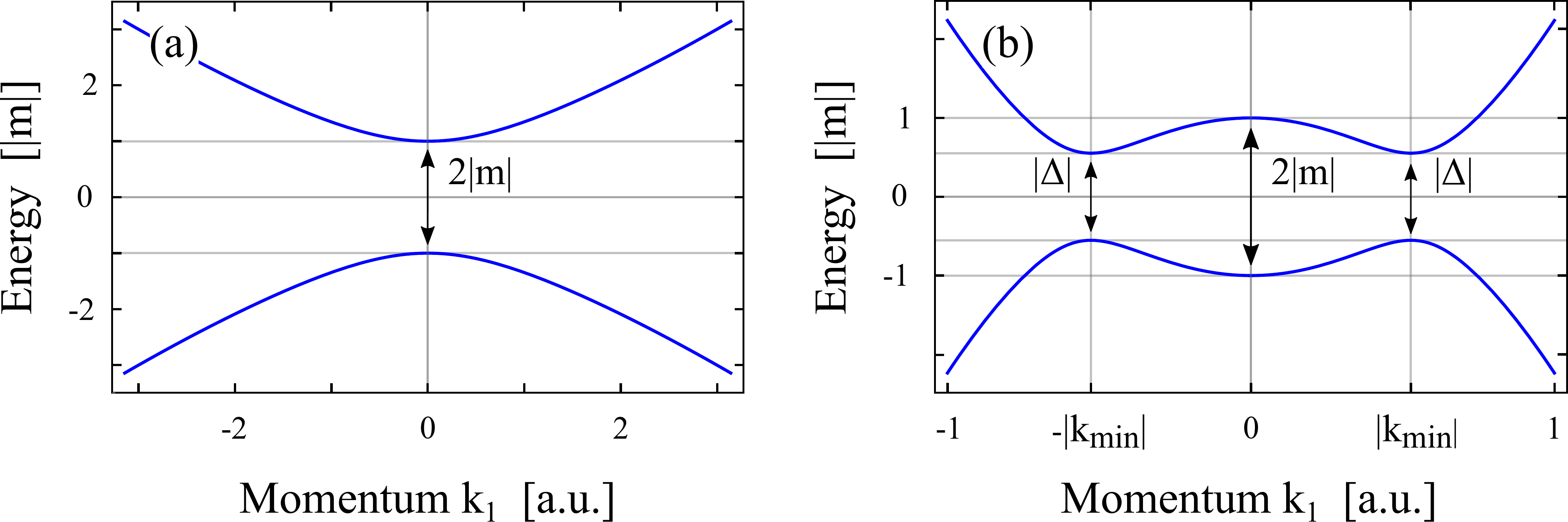}
\caption{Band structure corresponding to a single (pseudo-) spin block of the BHZ model for $k_2=D=0$ and for \textbf{(a)} a topologically non-trivial phase with $m\!=\!A\!=\!1$ and $B\!=\!0.1$. The minimal gap $2 \vert m \vert$ is located at the $\Gamma$-point.
\textbf{(b)} For a topologically non-trivial phase with $m\!=\!A\!=\!1$ and $B\!=\!3$. The minimal gap $\vert \Delta \vert$ is located at $\vert \textbf{k} \vert=\pm \vert k_\mathrm{min}\vert$.}
\label{Spectrum}
\end{figure}

Two-dimensional QAH insulators like (Hg,Mn)Te quantum wells or magnetically doped (Bi,Sb)Te thin films can be described by the Bernevig-Hughes-Zhang (BHZ) Hamiltonian consisting of the  two  \mbox{(pseudo-)spin} blocks $\mathcal{H}_\pm ( \mathbf{k} )$ \cite{Bernevig06,Note15,Note16}:
\footnotetext[15]{For the scope of this work, bulk inversion asymmetry terms are unimportant. Therefore, they are neglected throughout the manuscript.}
\footnotetext[16]{The basis of the BHZ Hamiltonian is system dependent. For (Hg,Mn)Te quantum wells,  it is for instance  ($\vert E_1, + \rangle$,  $\vert H_1, + \rangle$, $\vert E_1, - \rangle$,  $\vert H_1, - \rangle$).}
\begin{align} \label{BHZ}
\mathcal{H}_\mathrm{BHZ}(k)= \begin{pmatrix}
\mathcal{H}_+ ( \mathbf{k} ) & 0 \\
0 & \mathcal{H}_-^\star ( \mathbf{-k} ) 
\end{pmatrix}  \ .
\end{align}
On its own, each (pseudo-)spin block corresponds to a single Chern insulator in 2+1 space-time dimensions with
\begin{align} \label{hamchern}
\! \! \! \mathcal{H}_\pm ( \mathbf{k} ) \! = \! \pm \left(m_\pm \! - \! B \vert \mathbf{k} \vert^2\right) \! \sigma_\mathrm{3}  -  D \vert \mathbf{k} \vert^2 \sigma_{0}  +  A\left(k_\mathrm{1} \sigma_\mathrm{1}\!-\! k_\mathrm{2} \sigma_\mathrm{2} \right). 
\end{align}
As such, a single (pseudo-)spin block of the BHZ model is directly related to the parity anomaly of QED$_{2+1}$. In Eq.~\eqref{hamchern}, $\vert \mathbf{k} \vert^2  \! = \!  k_\mathrm{1}^2\! + \! k_\mathrm{2}^2$, $\sigma_{1,2,3}$ represent the Pauli matrices, $A$ is proportional to the Fermi velocity, and $D$ encodes a particle-hole asymmetry. The parity-breaking mass terms are $m_\pm$ and  $B\vert \mathbf{k} \vert^2$. 
In the QAH phase, only one of the two (pseudo-)spin blocks is topologically non-trivial with the finite Chern number \cite{Lu10,Qi08} 
\begin{align} \label{Chernnumber}
\mathcal{C}_\mathrm{QAH}^i= i \left( \mathrm{sgn}(m_i)+\mathrm{sgn}(B) \right)/2 \ ,   \quad i \in \lbrace +,- \rbrace \ .
\end{align}
The trivial (pseudo-)spin block has zero Chern number. Therefore, the entire topological response in the QAH phase is captured by a single Chern insulator in 2+1 space-time dimensions. If not stated otherwise, we focus on such a system in the remaining part of this work. In particular, we neglect the (pseudo-)spin subindex $\pm$. Moreover, we consider particle-hole symmetric Chern insulators since the $D\vert \mathbf{k} \vert^2$ term in Eq.~\eqref{hamchern} is parity-even and thus does not contribute to the parity anomaly \cite{JanPaper}.\\
The spectrum associated to Eq.~\eqref{hamchern} for $D=0$ is given by 
\begin{align} \label{bulkspectrum}
\epsilon_\mathrm{\pm}= \pm \sqrt{A^2 \vert \mathbf{k} \vert^2+(m-B \vert \mathbf{k} \vert^2)^2} \ .
\end{align}
Here, $\pm$ encodes the conduction and the valence band, respectively. In Fig.~\ref{Spectrum},  we show the influence of the mass parameters on the band structure. While the Dirac mass $m$  defines the mass gap at the $\Gamma-$point, the momentum dependent $B\vert \mathbf{k} \vert^2$ term acts like an effective mass  of a non-relativistic fermion system. Depending on the values for $m$, $B$, and $A$, the band structure significantly changes. For $m/B>0$, the system is topologically non-trivial with $\mathcal{C}_\mathrm{QAH}=\pm 1$. However, depending on the absolute values of the input parameters, the minimal gap can either be located at the $\Gamma$-point [Fig.~\ref{Spectrum}(a)], driven by the Dirac mass alone,  or apart from $\vert\mathbf{k}\vert=0$ [Fig.~\ref{Spectrum}(b)] at $k_\mathrm{min}\!=\!\pm \sqrt{2 m B-A^2}/(\sqrt{2}B)$.
The minimal gap  is therefore either defined by $2\vert m \vert$, or by the absolute value of
\begin{align}
\Delta=  A \sqrt{4 m B-A^2}/B  \ .
\end{align}
In contrast, for $m/B<0$, the system is topologically trivial,  characterized by $\mathcal{C}_\mathrm{QAH}= 0$. In this case  the minimal gap is  always located at the $\Gamma$-point.
At $m=0$, the topological phase transition occurs, which  comes along with a gap closing at $\mathbf{k} \! = \! 0$.
While in our plots we mostly consider positive mass terms, the inverted Dirac mass of an experimental QAH insulator is negative. Changing the overall sign of $m$ and $B$ alters the sign of the Chern number, but not the underlying physics.
Notice, that according to Eq.~\eqref{bulkspectrum} the Newtonian mass term does not make the spectrum bounded. It is from this perspective expected that the
$B \vert \mathbf{k} \vert^2$ term does not render the effective action of a Chern insulator UV finite.

Having discussed the particular influence of the mass terms on the band structure, let us emphasize one more time that both of them explicitly break the  parity symmetry in 2+1 space-time dimensions, defined as invariance of the theory under \mbox{$\mathcal{P}: \, (x_0,x_1,x_2) \rightarrow (x_0,-x_1,x_2)$}. Hence, the Dirac and the Newtonian mass term are directly related to the parity anomaly of massless QED$_{2+1}$. To concretize this statement, let us briefly review the concept of quantum anomalies.

\vspace{-.2cm}

\section{Parity Anomaly} \label{redlich}

\vspace{-.2cm}

In what follows, we consider Chern insulators in 2+1 space-time dimensions coupled to an external fluctuating abelian U(1) gauge field $\mathcal{A}_\mu$ [cf.~Eq.~\eqref{hamchern}]. Such models can be quantized by calculating the partition function $\mathcal{Z}[\mathcal{A}]$ via integrating out the fermionic degrees of freedom
\begin{align} \label{eq:partition}
\mathcal{Z}[\mathcal{A}]= \dfrac{1}{\mathcal{Z}[0]}  \int \! \mathrm{d} \bar{\psi} \, \mathrm{d} \psi \  \mathrm{e}^{\im \mathcal{S}[\bar{\psi},\psi,\mathcal{A}]}= \dfrac{1}{\mathcal{Z}[0]}  \, \mathrm{e}^{\im  S_\mathrm{eff}[\mathcal{A}]} \, . 
\end{align}
Here, $\psi$ and $\bar{\psi}$ are the two-component Dirac spinor and its adjoint. The bare and the effective action are defined by $\mathcal{S} [\bar{\psi},\psi,\mathcal{A}]$ and $\mathcal{S}_\mathrm{eff}[\mathcal{A}]$. If $\mathcal{S}_\mathrm{eff}[\mathcal{A}]$ has less symmetries than $S[\bar{\psi},\psi,\mathcal{A}]$, a quantum anomaly is present. For instance, massless QED$_{2+1}$ is described by the parity-symmetric bare action
\begin{align} \label{ActionQED}
\mathcal{S}[\bar{\psi},\psi,\mathcal{A}]= \int \! \! \mathrm{d}^3 x \ \im  \bar{\psi}  \left( \slashed \partial + \im e \slashed{\mathcal{A}} \right) \psi \, .
\end{align}
Here, we use the Feynman slash notation $\slashed \partial = \gamma_\mu \partial^\mu$ with the 2+1 dimensional Dirac matrices $\gamma_\mu$=($\sigma_3$,$\im \sigma_2$,$\im \sigma_1)$, and consider the metric tensor $g^{\mu \nu}\!=\!\mathrm{diag(+,-,-)}$. The associated effective action is given by 
\begin{align} \label{Seff}
\mathcal{S}_\mathrm{eff}[\mathcal{A}]= - \im \, \mathrm{Ln} \, \mathrm{det}\left[ (\slashed \partial + \im e \slashed{\mathcal{A}})/\slashed \partial \right] \ . 
\end{align}
As it was shown initially in Ref.~[\citen{Redlich83}],
the fermion determinant in Eq.~\eqref{Seff} changes sign under large gauge transformations\footnote{For U(1) gauge fields such transformations only exist on compact non-trivial manifolds, such as $\mathcal{M}=T^2 \times S_1$.} of odd winding $\omega=2 n \! + \! 1$ with $n \! \in\!  \mathbb{Z}$
\begin{align} \label{fermiondet}
 \mathrm{det}\left[ (\slashed \partial +  \im e \slashed{\mathcal{A}})/\slashed \partial \right] & \rightarrow (-1)^{\vert \omega \vert} \  \mathrm{det}\left[ (\slashed \partial + \im e\slashed{\mathcal{A}})/\slashed \partial \right] \\
 \Rightarrow \quad \mathcal{S}_\mathrm{eff} & \rightarrow \mathcal{S}_\mathrm{eff} \pm \pi \vert \omega \vert \ . \nonumber
\end{align}
Hence, as it stands $\mathcal{S}_\mathrm{eff}[\mathcal{A}]$ is not a well-defined object. This can also be seen by calculating the fermion determinant explicitly, which leads to a divergent expression. Consequently, the theory requires  a regularization scheme which needs to be chosen such that it ensures infinitesimal- as well as  large gauge invariance of $\mathcal{Z}[\mathcal{A}]$. In particular, it was shown  that each regularization scheme needs to break parity symmetry to ensure  gauge invariance. This is known as the parity anomaly \cite{Niemi83,Redlich83}.
The parity anomaly therefore  results from the particular regularization of QED$_{2+1}$. 
Even though a massive  QED$_{2+1}$ system breaks parity on the classical level, the fermion determinant still diverges and lacks gauge invariance. Again, this requires a parity-breaking regularization scheme, which ensures an integer quantized DC Hall conductivity associated to a gauge invariant Chern-Simons term in the effective action.
This property extends the peculiarities of the parity anomaly to the massive case \cite{Niemi83,Redlich83,Dunne98,schaposnik96}.
In this work, we show that adding the Newtonian mass term to a massive QED$_{2+1}$ Lagrangian also leads to an integer quantized DC Hall conductivity associated to  a gauge invariant Chern-Simons term. From this perspective the Newtonian mass term, provided by the material, acts similar to a parity-breaking element of a mathematical regularization scheme \cite{Note13}. Its contribution to the Chern number does not vanish in the parity-symmetric limit $m,B \rightarrow 0$.
Consequently, beside the Dirac mass also the Newtonian mass term is directly related to the parity anomaly. Inspired by this property, we compare in the following the role of the Newtonian mass term for the calculation of the effective action of a Chern insulator to common parity-breaking  regularization schemes of QED$_{2+1}$.

\vspace{-.2cm}

\section{Effective Action}

\vspace{-.2cm}

In what follows, we perturbatively evaluate the effective action corresponding to a single Chern insulator. Therefore, we Taylor expand the fermion determinant in Eq.~\eqref{fermiondet} to second order in the external background field $\mathcal{A}_\mu$. The free Lagrangian associated to Eq.~\eqref{hamchern} can be obtained by a Legendre transformation.  For $A \! =\! \hbar \! = \! 1$ and $D \! = \! 0$, this Lagragian is 
equivalent to a pure QED$_{2+1}$ Lagrangian except for an additional correction, which is quadratic in spatial derivatives
\begin{align} \label{freeLagrangian}
\mathcal{L}_0 & =\bar{\psi} \left( \im \slashed \partial -m \right) \psi - B \gamma^0 (\partial_i \psi )^\dagger (\partial^i \psi)  \ .
\end{align}
Here and in the following, we use  
the properties of the Dirac matrices given in App.~\ref{diracmatrices}. 
Coupling $\mathcal{L}_0$ covariantly to the U(1) gauge field $\mathcal{A}_\mu$, leads to the  Lagrangian 
\begin{align} \label{lagrangianQAH}
\!  \mathcal{L} & =\bar{\psi} \left( \im \slashed  D -m \right) \psi - B \gamma^0 (D_i \psi )^\dagger (D^i \psi) 
\\
&= \mathcal{L}_0  -   e \bar{\psi} ( \slashed{\mathcal{A}} \! + \!  e B \mathcal{A}^i \! \mathcal{A}_i ) \psi +  \im e B   \mathcal{A}^i \! \left(  \bar \psi (\partial_i \psi) \! - \!  (\partial_i \bar \psi) \psi   \right) 
\! , \nonumber
\end{align}
where $D_\mu= \partial_\mu + \im \mathrm{e} \mathcal{A}_\mu$ is the covariant derivative. 
From the interaction terms in  $\mathcal{L}$, we can read off the vertex contributions. For an incoming electron of momentum $k$, incoming photons of momentum $p$, and an outgoing electron of momentum $k \!+\! p$, we find the vertices
\begin{align}
V_{(1)}^\mu (2k \! + \! p) & = - \im e (\gamma^\mu - B \, \delta^{\mu}_{ \ \, i} \,  (2k \! + \! p)^i ) \label{linvertex} \\
V_{(2)}^{\mu \nu} (k) & = - 2 \im e^2 B \, \delta^{\mu}_i \delta^{ i \nu } \ . \label{vertexquad}
\end{align}
\noindent
Here, the subscript defines the number of involved photons.
In comparison to pure QED$_{2+1}$, the extra $B\vert \mathbf{k} \vert^2$ term in the Lagrangian  renormalizes the gauge-matter coupling. The original QED vertex in Eq.~\eqref{linvertex} obtains a momentum dependent correction. Additionally, Eq.~\eqref{vertexquad} defines a new vertex structure, which is of second order in gauge fields. This vertex encodes the diamagnetic response of the Chern insulator in Eq.~\eqref{hamchern} \cite{Czycholl}. 
The fermion propagator associated to the Lagrangian  $\mathcal{L}$ is given by 
\begin{figure}[t]
\centering
\includegraphics[scale=0.45]{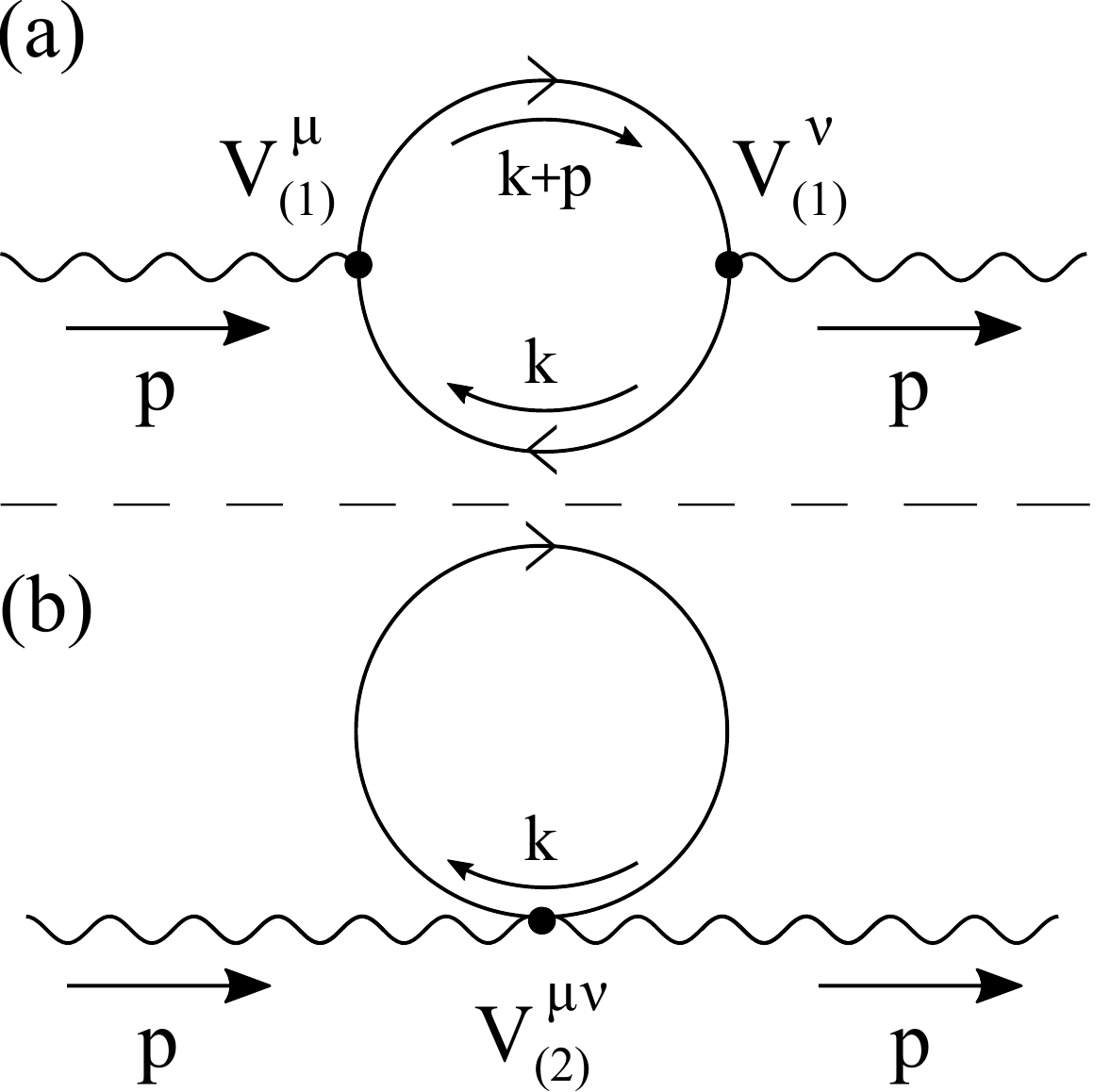}
\caption{Feynman diagrams for the vacuum polarization operator of a Chern insulator with vertices $V_{(1)}^{\mu,\nu}$ and $V_{(2)}^{\mu \nu}$. External momenta are denoted by $p$, loop momenta by $k$.}
\label{togloop}
\end{figure}
\begin{align}
S(k)= \dfrac{\im}{\slashed k - (m \! - \! B \vert \mathbf{k} \vert^2) + \im \epsilon }= \im \dfrac{\slashed k + M(k)}{k^2 - M(k)^2 + \im \epsilon} \ .
\end{align}
Here, we defined the momentum dependent mass term $M(k)= m - B \vert \mathbf{k} \vert^2$ and used the Feynman prescription with $\epsilon \! \rightarrow \! 0^+$.
To perturbatively obtain the second order effective action in external fields, we need to calculate the vacuum polarization operator $\Pi^{\mu \nu}(p)$ \footnote{Since we only consider zero chemical potential, the linear term in $\mathcal{A}_\mu$  needs to vanish. Physically, this term calculates the density, which is zero in the mass gap if there is no underlying  magnetic field.},
\begin{align}
\mathcal{S}_\mathrm{eff} = \dfrac{1}{2} \int \dfrac{\mathrm{d}^3 p}{(2 \pi)^3} \ \mathcal{A}_\mu(-p) \Pi^{\mu \nu}(p)  \mathcal{A}_\nu(p) \ .
\end{align}
Due to the vertex structure in Eqs.~\eqref{linvertex} and~\eqref{vertexquad}, the vacuum polarization operator is obtained by the sum of two \mbox{one-loop} Feynman integrals
\begin{align} \label{PObk2}
& \im \Pi^{\mu \nu} (p) =\im \Pi^{\mu \nu}_\mathrm{2a} (p)+\im \Pi^{\mu \nu}_\mathrm{2b} (p) \\ &=  - \! \! \int \! \frac{\mathrm{d}^3 k}{(2 \pi)^3} \, \mathrm{Tr}\big[S(k)V^\mu_{(1)}(2k\!+\!p)S(k\!+\!p)V^\nu_{(1)}(2k\!+\!p)  \nonumber\\
& \qquad \qquad \quad \ \ \, + S(k)V^{\mu \nu}_{(2)}(k)\big] \ , \nonumber
\end{align}
which are diagrammatically illustrated in Fig.~\ref{togloop}.
We start with the calculation of the first term in Eq.~\eqref{PObk2}, $\im\Pi^{\mu \nu}_\mathrm{2a}(p)$, which is the usual QED$_{2+1}$ vacuum polarization operator with renormalized vertex and propagator structure. As shown in Fig.~\ref{togloop}(a), this tensor is given by
\begin{widetext}
\begin{align} \label{poexp}
\im \Pi^{\mu \nu}_\mathrm{2a} (p)= \dfrac{e^2}{(2 \pi)^3} \int \! \mathrm{d}^3 k \ \dfrac{\mathrm{Tr} \left ([\gamma^\mu - B \, \delta^{\mu}_{\ i} (2k \! + \! p)^i \, \sigma_0] \im [\slashed k + M(k)] [ \gamma^\nu  - B \, \delta^{\nu}_{\ j} (2k \! + \! p)^j \, \sigma_0] \im [(\slashed k + \slashed p) + M(k \! + \! p)] \right) }{(k^2 - M(k)^2 + \im \epsilon)((k \! + \! p)^2 - M(k \! + \! p)^2 + \im \epsilon)} \ .
\end{align}
\end{widetext}
There are four different contributions to the Dirac trace $\mathrm{Tr} = \mathrm{Tr}_{\gamma \gamma} + \mathrm{Tr}_{\gamma 0} + \mathrm{Tr}_{0 \gamma}+ \mathrm{Tr}_{00}$, where the subscript defines the Dirac and identity part of the vertex structure in Eq.~\eqref{linvertex}. 
Since all physical response functions are given as functional derivatives of the effective action at zero external spatial momentum, from now on we focus on the calculation of 
$\im \Pi^{\mu \nu} (p_0,\mathbf{p}\! = \!0$). With this assumption, $p^2=p_0^2$, $\epsilon^{\mu \nu \lambda} p_\lambda= \epsilon^{\mu \nu 0} p_0 $, and  $M(k)=M(k \! + \! p)$.
Next, we introduce the Feynman parameter $x \in [0,1]$ and shift the loop momentum according to $k \!=\! l \!- \!px $. This gives the denominator in Eq.~\eqref{poexp} a quadratic form, 
allowing us to drop all linear terms in $l$ in the numerator, due to an anti-symmetric integration over symmetric boundaries \cite{peskin}. With $\alpha=\vert \mathbf{l} \vert^2$, this leads to [cf.~App.~\ref{diracmatrices}]
\begin{align} \label{trace}
\mathrm{Tr}& = \! 2 g^{\mu \nu} l^2 -4 l^\mu l^\nu -2x(1 \! - \! x)p_0^2  \\ \nonumber
& -\! 2g^{\mu \nu} \! \left[ M(\alpha)^2 \!+\! x(1 \! - \! x)p_0^2 \right]  \!-\! 2  \im \! \left[ M(\alpha)+ 2  B \alpha \right] \epsilon^{\mu \nu 0}  p_0  \\
 & + \! 4 B \alpha \delta^{\nu}_{m} \delta_m^{ \mu }   \!  \left[ 2 M(\alpha)    \!  - \!  B    \! \left[l_0^2 \!-\! \alpha \!-\! x(1 \! - \! x) p_0^2  \! + \! M(\alpha)^2 \right]  \right] \! . \nonumber 
\end{align}
In the QED$_{2+1}$ limit $B \! \rightarrow \! 0 $, the Dirac trace in Eq.~\eqref{trace} reduces to the well-known result \cite{peskin,ashok,Schwartz} 
\begin{align} \label{QEDlimit}
\mathrm{Tr} &= 2 g^{\mu \nu} l^2 - 4 l^\mu l^\nu  +4 x(1 \! - \! x)p_0^2 \\ & - 2g^{\mu \nu} \left( m^2 +x(1 \! - \! x)p_0^2 \right) - 2 \im  m \epsilon^{\mu \nu 0} p_0  \ . \nonumber
\end{align}
Notice, that the off-diagonal Chern-Simons contribution in Eq.~\eqref{QEDlimit} gets shifted by the renormalized vertex structure in Eq.~\eqref{trace}. 
As argued above, this will lead to an integer quantized DC Hall conductivity associated to a gauge invariant Chern-Simons term. 

\subsection{Off-Diagonal Response}

\noindent
To prove this statement, we  evaluate the integral
\begin{align} \label{csint}
&\im \Pi_\mathrm{CS}^{\mu \nu}(p_0,\mathbf{p}=0)  \\
&= \dfrac{ e^2}{(2 \pi)^3} \int \limits_0^1 \! \! \mathrm{d}x \int  \! \!  \mathrm{d}^3 l \, \dfrac{-2  \im \epsilon^{\mu \nu 0}  \left( M(\alpha) \! + \! 2 B \alpha  \right) p^0}{(l_0^2 -\alpha -M(\alpha)^2+x(1 \! - \! x)p_0^2 + \im \epsilon)^2}      \nonumber \\ & = \dfrac{  e^2}{8 \pi} \epsilon^{\mu \nu 0}  p_0 \int \limits_0^1 \! \! \mathrm{d}x \! \int  \limits_0^\infty \! \!  \mathrm{d} \alpha \, \dfrac{ m +  B \alpha   }{(\alpha + M(\alpha)^2-x(1 \! - \! x)p_0^2 - \im \epsilon)^{3/2} }  . \nonumber
\end{align}
Here, we used the Feynman parametrization and solved the complex time-integration via the residue theorem.
For Chern insulators, time and spatial momenta need to be integrated separately since the $B \vert \mathbf{k} \vert^2$ term breaks the Lorentz symmetry. Hence, it is not possible to Wick-rotate and integrate over an Euclidean three-sphere. Integrating the Feynman parameter $x$ implies
\begin{align}
& \int_0^1 \! \mathrm{d}x \, \dfrac{1}{(\alpha+M(\alpha)^2-x(1 \! - \! x)p_0^2-\im \epsilon)^{3/2}} \\
& \quad \quad \quad= \dfrac{4}{\sqrt{\alpha+M(\alpha)^2-\im \epsilon} \, (4\alpha+4 M(\alpha)^2-p_0^2-4\im \epsilon)} \nonumber  \ ,
\end{align}
where we kept the $\im \epsilon$-prescription to circumvent the poles for $\alpha\!>\!0$, appearing if $p_0$ exceeds the gap.
Finally, we perform the remaining $\alpha$-integration for an arbitrary driving frequency $p_0$ and subsequently set $\epsilon \rightarrow 0^+$. Due to its lengthy form, we present the general Chern-Simons contribution and its AC Hall conductivity $\sigma_\mathrm{xy}(p_0)$ in App.~\ref{fullsol},
\begin{align}
\im \Pi_\mathrm{CS}^{\mu \nu}(p_0,\mathbf{p}=0) &= \sigma_\mathrm{xy}(p_0) \epsilon^{\mu \nu 0}  p_0 \ .
\end{align}
Instead, let us first analyze the  Taylor expansion of the AC Hall conductivity in terms of the frequency $p_0$,
\begin{align} \label{taylorexp}
\sigma_\mathrm{xy}(p_0) & = \sigma_\mathrm{xy}(0)+ \frac{ \sigma_\mathrm{xy}''(p_0) \vert_{p_0=0}}{2!} p_0^2  + \mathcal{O}(p_0^4)
\end{align}
with the coefficients (reintroducing $A$ and $\hbar$)
\begin{align}
\sigma_\mathrm{xy}(0) &= \dfrac{e^2 }{2 h}  \left[ \mathrm{sgn}(m) \!+ \! \mathrm{sgn}(B) \right] = \dfrac{e^2}{h} \, \mathcal{C}_\mathrm{QAH} \, , \label{DCcondcutivity}\\
\sigma_\mathrm{xy}''(p_0)\vert_{p_0=0} &=\dfrac{  e^2 }{ h}  \left[ \dfrac{2 \, \mathcal{C}_\mathrm{QAH}}{3 \Delta^2}    -  \dfrac{A^4 }{\Delta^2 B^2} \frac{\mathrm{sgn}(m) }{24 m^2 }  \right] \ . \label{ACcorrection}
\end{align}
\noindent
Equation \eqref{DCcondcutivity} defines the DC Hall conductivity of  the single Chern insulator given in Eq.~\eqref{hamchern}. In the QAH phase, this value matches the DC Hall conductivity of the entire system, as only one Chern insulator contributes to the topological response \cite{Lu10,Qi08}.
In comparison to a pure QED$_{2+1}$ system with a half-quantized $\sigma_\mathrm{xy}(0)$, the Newtonian mass term ensures integer quantization. Hence, the associated Chern-Simons term is  gauge invariant \footnote{Notice, that the full renormalized effective action of QED$_{2+1}$ is gauge invariant due to the presence of non-analytic terms which cannot be derived perturbatively \cite{Narayanan97}}.

In contrast, Eq.~\eqref{ACcorrection} defines the leading order AC correction to $\sigma_\mathrm{xy}(0)$, which contains two terms of different  origin. One the one hand, there is a term proportional to the Chern number $\mathcal{C}_\mathrm{QAH}$. 
In the trivial phase $m/B\!<\!0$, this term vanishes and the first order AC correction is solely given by the second term in Eq.~\eqref{ACcorrection}.
Instead, for $m/B>0$, this term contributes to the first order AC correction. In experimental systems like (Hg,Mn)Te quantum wells with $m_+\!=\!-10$meV, $B\!=\!-1075 \mathrm{meV}\mathrm{nm}^2$, and $A\!=\!365 \mathrm{meV}\mathrm{nm}$, the term proportional to $\mathcal{C}_\mathrm{QAH}$ defines $\approx \! 10\%$ of the entire signal in Eq.~\eqref{ACcorrection}. From a theoretical point of view, this term is induced by a finite Newtonian mass, which breaks the Lorentz symmetry in Eq.~\eqref{lagrangianQAH}. Consequently, it vanishes in 
 the QED limit $B \! \rightarrow \! 0 $, since 
\begin{align}
\lim_{B \rightarrow 0} \, \dfrac{1}{\Delta^2} =0 \quad \wedge \quad 
\lim_{B \rightarrow 0} \, \dfrac{A^4 }{\Delta^2 B^2}= - 1 \ .
\end{align}
In this limit, Eq.~\eqref{ACcorrection} reduces to the QED result
\begin{align}
\lim_{B \rightarrow 0}
\sigma_\mathrm{xy}''(p_0)\vert_{p_0=0} &=     \dfrac{  e^2 }{\mathrm{h} }   \frac{\mathrm{sgn}(m)  }{24 m^2 }  \ .
\end{align}
Due to its unique relation to the Newtonian mass, the term proportional to the Chern number in Eq.~\eqref{ACcorrection} is quadratically suppressed by the ratio of $p_0$ over the gap $\vert \Delta \vert$. In contrast, the second term in Eq.~\eqref{ACcorrection}, which is the first order QED correction to the DC Hall conductivity, is quadratically suppressed by $p_0$  over the Dirac mass. According to the quadratic suppression of both terms, the AC signal in Eq.~\eqref{ACcorrection}
stays close to quantized values and 
matches the AC Hall response of the entire QAH system for small driving frequencies $p_0$.

Let us briefly comment on how to experimentally disentangle the QED from the Newtonian part in the first order AC Hall correction.
In QAH insulators like (Hg,Mn)Te quantum wells the topological phase transition originates from a sign change of the Dirac mass of one of the two Chern insulators. In what follows, let us assume it is $m_+$, meaning that the topological phase transition takes place in the (pseudo-)spin up block of the BHZ model [cf.~Eq.~\eqref{BHZ}]. Due to the parameters above, this  transition is associated to an overall sign change of $\sigma_\mathrm{xy}''(p_0)\vert_{p_0=0}$, which is mainly driven by the QED correction in Eq.~\eqref{ACcorrection}. Consequently, measuring the AC Hall signal at $\pm m_+$ \footnote{Throughout this experiment, $m_-$ needs to be fixed in the topologically trivial region.}, allows to substract the QED correction and, therefore, to isolate the  contribution to Eq.~\eqref{ACcorrection} which is induced by the Newtonian mass.
Moreover, $\sigma_\mathrm{xy}(p_0)$ is related to the Faraday and the Kerr angle of two-dimensional QAH insulators\cite{Allan10,Allan11}. The first term in Eq.~\eqref{ACcorrection} can be therefore resolved by magneto-optical experiments, as well. Let us consider a linearly polarized  electric field, which incidents normally on the QAH system. For frequencies much smaller than the gap, which justify Eq.~\eqref{taylorexp}, one finds \cite{Allan11} 
\begin{align}
\Theta_\mathrm{F}(p_0)  &=  \ \, \; \mathrm{Arctan} \left[ \dfrac{ \pi \sigma_\mathrm{xy}(p_0)}{ \epsilon_0 \mathrm{c}} \right] \\
\Theta_\mathrm{K}(p_0)  &= - \mathrm{Arctan} \left[ \dfrac{ \epsilon_0 \mathrm{c}}{ \pi \sigma_\mathrm{xy}(p_0)} \right] \ .
\end{align}
Here, $\mathrm{c}$ is the vacuum speed of light and $\epsilon_0$ the vacuum permittivity. While these identities imply  quantized values of the Faraday and Kerr angles in the DC limit, they carry the information of how these angles  change due to the contribution of the first term in Eq.~\eqref{ACcorrection}. To resolve this effect in one of these experiments, $\vert p_0 \vert \! \ll \! \mathrm{Min}(2 \vert m \vert ,\vert \Delta \vert)$. For instance in inverted (Hg,Mn)Te/CdTe quantum wells, the gap is of the order of several meV
\cite{Beugeling2012}, depending on the particular manganese concentration. This corresponds to 
frequencies in the THz regime.
For such frequencies, the first order correction to the DC Faraday and Kerr angles is on the order of milli-rad, which can be resolved by recent Faraday polarimeters \cite{Morris12}.

Before we discuss the general solution of the AC Hall conductivity, note that the non-quantized value for  $\sigma_\mathrm{xy}(p_0 \! \neq \! 0)$ makes the associated Chern-Simons term (large) gauge non-invariant. Since analogously 
to thermal effects, an AC driving field excites non-topological degrees of freedom, this effect
corresponds to the non-gauge invariance of finite temperature Chern-Simons terms. For these theories, it was shown that the full effective action contains non-perturbative corrections in $\mathcal{A}_\mu$, absorbing this non-invariance \cite{dunne97,schaposnik99}. These terms cannot be found by the Taylor expansion of the fermion determinant Eq.~\eqref{fermiondet}. However, due to the fact that they are higher order in gauge fields, they do not contribute to the conductivity.\\

\begin{figure}[t]
\centering
\includegraphics[scale=.2]{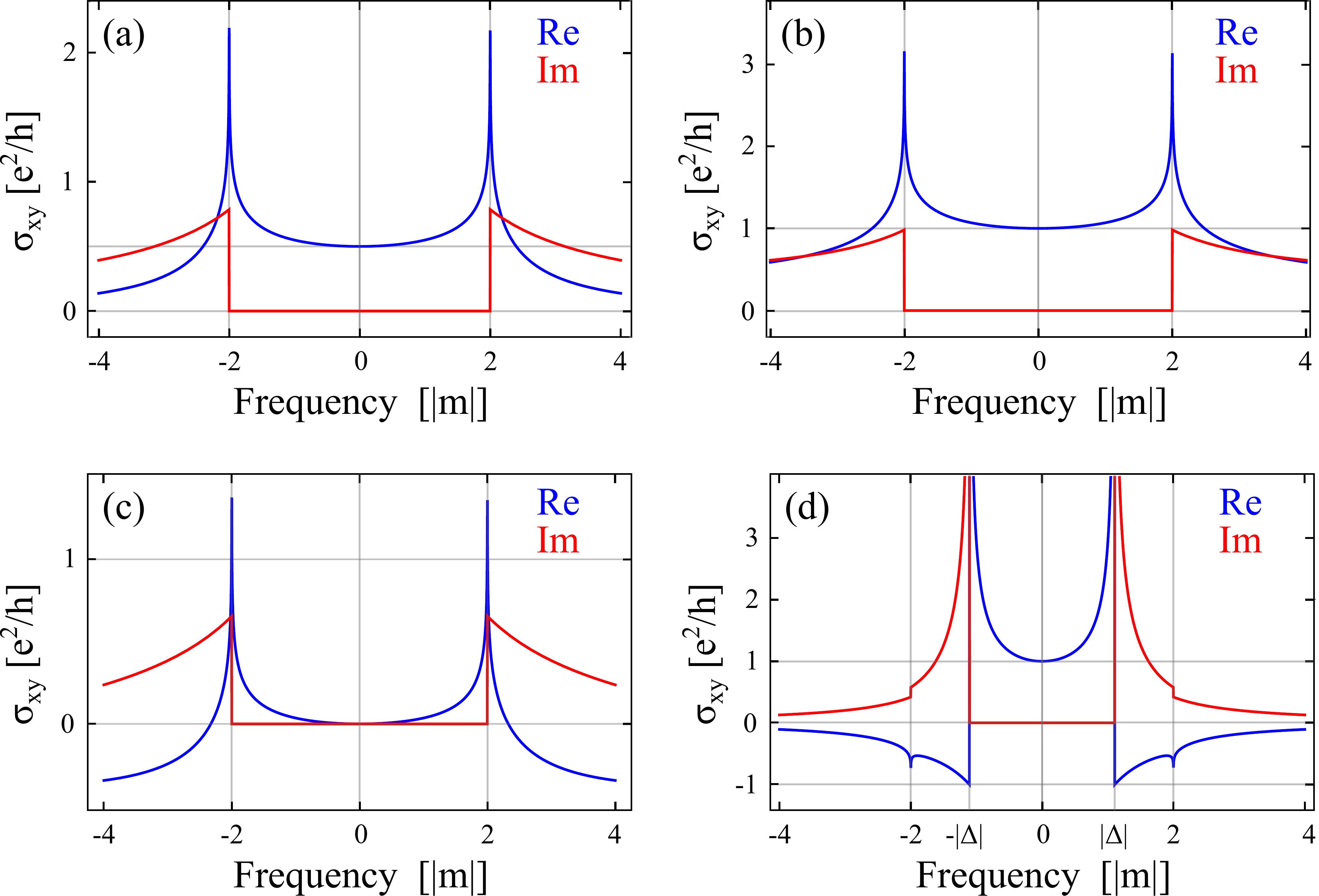}
\caption{
Hall conductivity $\sigma_\mathrm{xy}(p_0)$ for (\textbf{a}) a QED$_{2+1}$ system with $m\!=\!A\!=\!1$ and $B\!=\!0$, (\textbf{b}) a non-trivial Chern insulator with  $m\!=\!A\!=\!1$ and $B\!=\!0.1$, (\textbf{c}) a trivial Chern insulator with  $m\!=\!A\!=\!1$ and $B\!=\!-0.1$ and (\textbf{d}) a non-trivial Chern insulator with  $m\!=\!A\!=\!1$ and $B\!=\!3$. While the spectrum associated to (a)-(c) has the minimal gap $2 \vert m \vert$ at the $\Gamma$-point, the spectrum related to (d) has the minimal gap $\vert \Delta \vert$ at $k_\mathrm{min}$ [cf.~Sec.~\ref{model}]. Notice that all discontinuities/singularities arise from the assumption of zero temperature and disorder. Taking into account these ingredients makes all curves continuous.}
\label{reim}
\end{figure}

Let us now analyze the general solution of the AC Hall conductivity corresponding to a single Chern insulator. Figure~\ref{reim} shows the real and the imaginary part of 
$\sigma_\mathrm{xy}(p_0)$ according to its general form  in App.~\ref{fullsol}. To study the influence of the Newtonian mass term, Fig.~\ref{reim}(a) shows $\sigma_\mathrm{xy}(p_0)$ for a pure QED$_{2+1}$ system. At $p_0\!=\!0$, one observes the characteristic  half-quantization. Moreover, the real part of $\sigma_\mathrm{xy}(p_0)$ shows a resonance at $p_0=\pm 2\vert m \vert$ and tends to zero for larger frequencies.
For $\vert p_0 \vert \geq 2 \vert m \vert$, the AC field excites particle-hole pairs which can  propagate unhindered for $p_0 \! =\! \pm 2\vert m \vert$. This is the origin of the resonance \cite{Hill11,HIll13,Lutchyn09}. For large frequencies, the AC field dominates the gap which protects the topological phase. This leads to a vanishing AC Hall conductivity.
The imaginary part of $\sigma_\mathrm{xy}(p_0)$ satisfies the Kramers-Kronig relation
\begin{align}
\mathrm{Re} \, \mathrm{\sigma}_\mathrm{xy}(p_0)=\dfrac{1}{\pi} \mathrm{P} \int \! \mathrm{d} p_0' \, \dfrac{\mathrm{Im}\, \mathrm{\sigma}_\mathrm{xy}(p_0')}{p_0'-p_0} \ .
\end{align}
It is zero in the mass gap, becomes finite for $\vert p_0 \vert \geq 2  \vert m \vert$ and decreases afterwards. 
Since $\mathrm{Im} \, \sigma_\mathrm{xy}(p_0)$ results from interband absorptions \cite{Hill11,Allan10}, it is only  non-zero if the external frequency is able to excite a finite density of states.

Figure~\ref{reim}(b) and~\ref{reim}(c) show the corresponding plots for a non-trivial and a trivial Chern insulator with minimal gap size $ 2 \vert m \vert $, respectively. The AC Hall conductivity shows the same features as a pure Dirac system, except for the integer quantization of its DC Hall conductivity.  However, for a minimal gap apart from the $\Gamma$-point, the situation differs, as shown in Fig.~\ref{reim}(d). Here, the first resonance of the real part occurs at $p_0=\pm \vert \Delta \vert$. The $p_0=\pm 2 \vert m \vert$ resonance persists, but peaks in opposite direction since the density of states now decreases at  $ p_0  = \pm 2 \vert m \vert$. This property can also be seen in  Im$\, \sigma_\mathrm{xy}(p_0)$, which 
resolves the Van Hove singularity
at $p_0 =\pm \vert \Delta \vert$ and drops at $ p_0  = \pm 2 \vert m \vert$. Consequently, measuring the AC conductivity provides the information if the minimal gap is defined by the Dirac mass at the $\Gamma$-point, or rather by an interplay between the Dirac and the Newtonian mass apart from $\mathbf{k}=0$. 

Above, we have discussed the AC Hall response of a single Chern insulator which is either in the topologically trivial phase with $m \! \equiv \! m_-$, or in the topologically non-trivial phase with $m \! \equiv \! m_+$. The entire QAH response of the two (pseudo-)spin blocks $\mathcal{H}_\pm (\mathbf{k})$ in the BHZ model [cf.~Eq.~\eqref{BHZ}] corresponds to the superposition of both of these signals. In contrast to the DC Hall conductivity, the AC Hall conductivity contains corrections of non-topological origin. As such, also the topologically trivial (pseudo-)spin block of the BHZ model significantly contributes to the entire QAH response if the frequency $ \vert p_0 \vert$ is not much smaller than the trivial gap $2 \vert m_- \vert$.

\vspace{-.2cm}

\subsection{Diagonal Response}

\vspace{-.2cm}

Having discussed the off-diagonal response, we are still left with the calculation of the diagonal parts in Eq.~\eqref{trace}. Since our system is a bulk insulator, we physically expect that these terms vanish for $p=0$. The diagonal contributions can be calculated via the same techniques as used above. This leads to [cf.~App.~\ref{Intindentities}]
\begin{align} \label{respo}
& \im \Pi_\mathrm{D,2a}^{\mu \nu}(p=0)= \\ &
\dfrac{\im e^2 \delta^{\mu}_i \delta^{ i \nu } }{4 \pi \vert B \vert } \! \left[   1 \! - \! 2mB \! - \! 2 \vert m \vert \vert B \vert \! - \!  \mathrm{Ln} \left[ \frac{ 4 B^2 \Lambda}{1-2Bm+2 \vert m \vert \vert B \vert  }\right] \! \right] \! , \nonumber
\end{align}
where $\Lambda$ is a hard momentum cutoff in $\alpha=\vert \mathbf{l} \vert^2$. So far, we focused on the contributions of the first Feynman diagram in Fig.~\eqref{togloop}. Using analog techniques and the quadratic vertex in external fields, Eq.~\eqref{vertexquad}, 
the second Feynman diagram in Fig.~\eqref{togloop} yields [cf. App.~\ref{Intindentities}]
\begin{align} \label{seconddia}
\im \Pi_\mathrm{D,2b}^{\mu \nu}(p=0) = -\dfrac{\im e^2  \vert B \vert \Lambda \delta^{\mu}_i \delta^{ i \nu }   }{2 \pi}    -  \im \Pi_\mathrm{D,2a}^{\mu \nu}(p=0)  \ . 
\end{align}
This expression exactly cancels the finite and logarithmic divergent terms in Eq.~\eqref{respo}. Nevertheless, the full effective action still contains a term proportional to the UV cutoff $\Lambda$ which diverges during renormalization $\Lambda \rightarrow \infty$. However, as stated above, the diagonal contribution should be regularized/renormalized such that it vanishes for $p \! = \! 0$. Physically, such a renormalization corresponds to a proper definition 
of the particle density. The second term in Eq.~\eqref{PObk2}, which corresponds to the quadratic vertex in gauge fields Eq.~\eqref{vertexquad},
encodes the diamagnetic response of our system. This response is proportional to the particle density and as such needs to vanish in the gap. Due to the fact that we did not renormalize the Dirac sea contribution to the particle density, e.g. by anti-symmetrization, the divergence in Eq.~\eqref{seconddia} persists.\\

As a consequence of the  underlying parity anomaly each regularization scheme for a single Chern insulator does either break parity or gauge symmetry. The naive introduction of the hard momentum cutoff $\Lambda$ in Eq.~\eqref{respo} breaks the gauge symmetry. To preserve this symmetry, we should rather choose a parity-breaking regulator, such as a PV field with mass terms $M$ and $B$ \footnote{To decouple the PV field during renormalization we tune its mass gap to infinity, $M \rightarrow \infty$.}. While this field by construction ensures 
\begin{align}
\im \Pi_\mathrm{D,2a}^{\mu \nu}(p=0)+\im \Pi_\mathrm{D,2b}^{\mu \nu}(p=0)=0 \ ,
\end{align}
it contributes to the AC Hall response, as it breaks parity. For the entire system, this  can be resolved by introducing a second PV field for the regularization of the second \mbox{(pseudo-)}spin block of the BHZ model. If this field is constructed such that both PV fields are Kramers (time-reversal) partners, their combined contribution to the AC Hall conductivity of the entire QAH system vanishes.

\vspace{-.2cm}

\section{Newtonian Mass in the Context of QED$_{2+1}$ Regularization Schemes}

\vspace{-.2cm}

Albeit the Newtonian mass term in the Hamilton of a single Chern insulator [cf.~Eq.~\eqref{hamchern}] is a physical parameter provided by the material, before renormalization it acts similar to a parity-breaking regulator of a pure QED$_{2+1}$ system in the calculation of the effective action \cite{Note20}. In what follows, let us concretize this statement.

In the context of quantum field theories there are plenty of different regularization schemes, each breaking different symmetries. As discussed in Sec.~\ref{redlich}, the regularization scheme associated to an odd number of 2+1 dimensional Dirac fermions needs to break parity symmetry to ensure gauge invariance of the effective action. Manifestly parity breaking regularization schemes are for example PV regularization, lattice regularization with Wilson fermions, and $\zeta$-function regularization, which we briefly review in Appendix~\ref{reviewregschemes} \cite{schaposnik96,Narayanan97,Dunne98,Schaposnik2017}. 
All these schemes induce a
parity odd Chern-Simons term  in the effective action of a QED$_{2+1}$ system, directly proportional to the sign of the regularization parameter which breaks parity. For the schemes mentioned, this is the PV mass, the Wilson parameter, as well as the integration contour in the $\zeta$-function regularization.  
Together with the Chern-Simons contribution induced by the finite  Dirac mass $m$, this leads to an integer quantized DC Hall conductivity\cite{schaposnik96, Coste89}.

Adding the Newtonian mass term to a pure QED$_{2+1}$ system also ensures this property [cf.~Eq.~\eqref{DCcondcutivity}].
Even if one would remove the physical parameter $B \! \rightarrow \! 0$ in the end of the calculation, its contribution to the DC Hall conductivity persists and the Chern number stays integer quantized. From this perspective the $B \mathbf{k}^2$ term acts similar to a parity breaking element of a certain regularization scheme. To concretize what we mean with 'similar', let us compare the role of the Newtonian mass term in the calculation of the effective action to common parity-breaking regulators before renormalization \cite{Note20}\footnotetext[20]{During renormalization each mathematical regulator needs to be removed from the theory. This is in strong contrast to the Newtonian mass term which is a real parameter of the system. Therefore, any comparison between the $B \mathbf{k}^2$ term and a regulator needs to be done before renormalization.}.
Since the $B\vert \mathbf{k} \vert^2$ term provides a momentum dependent Dirac mass correction, it is natural to compare its role in the calculation of the effective action to regularization schemes which add terms of higher order derivatives to the bare Lagrangian. Such approaches are for example lattice regularization with Wilson fermions\cite{Coste89} [App.~\ref{lattice}] and higher derivative regularization\cite{Nagahama1986} [App.~\ref{higher}]. However, except for the property that these schemes yield also an integer quantized Chern number, there are several key differences to the $B\vert \mathbf{k} \vert^2$ term. By construction, each regularization needs to render the effective action  finite \cite{peskin}. As shown in Eq.~\eqref{seconddia}, adding the Newtonian mass to a massive Dirac Lagrangian does not exhibit this property. To reduce the superficial degree of divergence, the higher derivative regularization multiplies the entire non-interacting Dirac Lagrangian by $(1+\partial^2/M^2)$. This has two implications. In contrast to the $B\vert \mathbf{k} \vert^2$ term, it circumvents the vertex renormalizations in Eqs.~\eqref{linvertex}~and~\eqref{vertexquad}, but as a price manifestly breaks local gauge invariance\footnote{This property  is reestablished during renormalization.}. Moreover, by construction this  approach also regularizes the kinetic part of the Lagrangian.

In a lattice approach,  the inverse lattice spacing $a^{-1}$ makes the theory finite. To avoid fermion doubling and to break the parity, the lattice QED$_{2+1}$ Lagrangian comes along with an additional Wilson mass term $\propto \! s a  k^2$. Here, $s=\pm1$ is the Wilson parameter and $k$ is the lattice three-momentum\cite{Ho84,Ho85,Coste89}.
Clearly, the Wilson mass is directly related to the $B \vert \mathbf{k} \vert^2$ term.  However, by construction the Wilson mass is Lorentz invariant, while the Newtonian mass breaks this symmetry. Further, the Wilson mass vanishes during renormalization, $a \rightarrow 0$, which is not the case for the $B \vert \mathbf{k} \vert^2$ term since it is a material parameter.

\vspace{-.2cm}

\section{Summary and Outlook}

\vspace{-.2cm}

In this work, we connected the Newtonian mass term of a QAH insulator to the parity anomaly of QED$_{2+1}$. In particular, we showed that before renormalization the Newtonian mass term acts similar to a parity-breaking regulator in the calculation of the effective  action. Hence, this mass is directly related to the regularization of QED$_{2+1}$ and, as such, to the parity anomaly. More precisely, we showed that the Newtonian mass alone does not render the effective action UV finite, but ensures an integer quantized DC Hall conductivity. Before renormalization, it acts similar to a Wilson mass term, which avoids the fermion doubling in a lattice approach.

Moreover, we derived the 
AC Hall conductivity of a QAH insulator during the calculation of the effective action. We showed that the leading order AC correction to the DC Hall conductivity contains a term proportional to the Chern number. This term originates from the  Newtonian mass and can be measured by purely electrical means, or by determining the leading order frequency correction to the DC Faraday or Kerr angles of our system. Further, we revealed that the Newtonian mass  significantly changes the resonance structure of the AC Hall conductivity in comparison to  pure QED$_{2+1}$ systems.

The classification of three-dimensional topological insulators is also given by the interplay between their Dirac and Newtonian mass term. The effective action of those systems can be calculated by the parity-anomaly induced Chern-Simons term in $4+1$ space-time dimensions \cite{Qi08}. 
Consequently, a natural extension of our work would be to derive the
effective action of a $4\!+\!1$ dimensional QED system including both, a Dirac as well as a momentum-dependent term.



\begin{acknowledgments}
We thank  F.~A.~Schaposnik, F.~Wilczek, G.~W.~Semenoff, T.~Kießling, B.~Trauzettel and L.~W.~Molenkamp  for useful discussions.  We acknowledge financial support through the Deutsche Forschungsgemeinschaft (DFG, German Research Foundation), project id 258499086 - SFB 1170 'ToCoTronics', the ENB Graduate School on Topological Insulators, the Würzburg-Dresden Cluster of Excellence on Complexity and Topology in Quantum Matter - ct.qmat (EXC 2147, project id 39085490), and in part through the National Science Foundation NSF~PHY-1748958. E.~M.~Hankiewicz thanks the KITP institute for its hospitality.
\end{acknowledgments}

\appendix

\section{Calculation of the Effective Action} \label{diracmatrices}
\label{Intindentities}

For the calculation of the trace in Eq.~\eqref{poexp}, we used the  properties of the  2+1 dimensional Dirac matrices \cite{kimura94}: 
\begin{align} \label{traceid}
\mathrm{Tr}\left[ \gamma_\mu \right] & =0 \\
\lbrace \gamma_\mu , \gamma_\nu \rbrace & = 2 g_{\mu \nu} \nonumber \\ \nonumber
\mathrm{Tr}\left[ \gamma_\mu \gamma_\nu \right]  & = 2 g_\mathrm{\mu \nu}\\  \nonumber
 \mathrm{Tr}\left[ \gamma_\mu \gamma_\nu \gamma_\lambda \right] & =  2 \im \epsilon_{\mu \nu \lambda} \quad \mathrm{with} \quad \epsilon_{0 1 2}=-1 \\ \nonumber
\mathrm{Tr}\left[ \gamma_\mu \gamma_\nu \gamma_\lambda \gamma_\rho \right] &= 2(g_\mathrm{\mu \nu} g_\mathrm{\lambda \rho} - g_\mathrm{\mu \lambda} g_\mathrm{\nu \rho} + g_\mathrm{\mu \rho} g_\mathrm{\nu \lambda}) \ . \nonumber 
\end{align}
With  $C^\mu = - B \, \delta^{\mu}_{\ i} (2k \! + \! p)^i$, this in particular leads to
\begin{align}
-\dfrac{1}{2}\mathrm{Tr}_{\gamma \gamma} & =   2 k^\mu k^\nu  +  k^\mu p^\nu +  k^\nu p^\mu \\ \nonumber
& \quad + g^{\mu \nu} \left( M(k)M(k \! + \! p) - k^2 - kp \right) \nonumber  \\ \nonumber
& \quad  - \im  \epsilon^{\mu \nu \lambda} \left( \left[ M(k \! + \! p)-M(k) \right]k_\lambda - M(k) p_\lambda \right)  \\ \nonumber
-\dfrac{1}{2}\mathrm{Tr}_{\gamma 0} & =  C^\nu g^{\mu \lambda}  \left[  \left(  M(k)+M(k \! + \! p) \right) k_\lambda + M(k) p_\lambda \right]  \\ \nonumber
& \quad +  \im C^\nu \epsilon^{\mu \lambda \sigma} k_\lambda (k \! + \! p)_\sigma \\ \nonumber
-\dfrac{1}{2}\mathrm{Tr}_{0 \gamma} & =  C^\mu g^{\nu \lambda}  \left[  \left(  M(k)+M(k \! + \! p) \right) k_\lambda + M(k) p_\lambda \right] \\ \nonumber & \quad - \im C^\mu \epsilon^{\nu \lambda \sigma} k_\lambda (k \! + \! p)_\sigma \\ \nonumber
-\dfrac{1}{2}\mathrm{Tr}_{00} & = C^\mu C^\nu \left[ g^{\lambda \sigma} k_\lambda \left( k \! + \! p \right)_\sigma + M(k) M(k \! + \! p) \right].
\end{align}

\noindent
Moreover, for the calculation of $\im \Pi^{\mu \nu}_\text{Diag,2a}$ and $\im \Pi^{\mu \nu}_\text{Diag,2b}$, we used the following identities for the integration of the spatial loop momentum\cite{intcit}:
\begin{align}
& \int   \mathrm{d}x \, \dfrac{1}{(a  x^2+b   x + c )^{3/2}} = - \dfrac{2( b + 2 a x)}{(b^2 - 4 a c)\sqrt{c+x(b+ a x)}} \nonumber \\
& \int   \mathrm{d }x \, \dfrac{x }{(a  x^2 +b   x + c )^{3/2}}  = \dfrac{4 c+ 2 b x}{(b^2 - 4 a c)\sqrt{c+x(b+ a x)}} \nonumber
\\
& \int \mathrm{d}x \dfrac{x}{\sqrt{ax^2+bx+c}}   =\dfrac{1}{a} \sqrt{ax^2+bx+c}  \\ & \qquad \qquad \qquad \qquad \quad - \dfrac{b}{2a} \int  \mathrm{d}x \dfrac{1}{\sqrt{ax^2+bx+c}} \ , \nonumber \\ \nonumber
&\int \mathrm{d}x  \dfrac{1}{\sqrt{ax^2+bx+c}}   \\
& = \begin{cases}
\dfrac{1}{\sqrt{a}} \, \mathrm{Ln} \left \vert 2 \sqrt{a(ax^2\!+\!bx\!+\!c)} \!+\! 2 ax \!+\!b \right \vert  \quad \mathrm{for} \ a>0\\
- \dfrac{1}{\sqrt{-a}} \, \mathrm{Arcsin} \dfrac{2 a x + b}{\sqrt{b^2-4 a c}}  \quad  \quad \quad \quad \quad  \ \, \mathrm{for} \   a<0
\end{cases} \nonumber  \ .
\end{align}

\section{Regularization Schemes} \label{reviewregschemes}

\noindent
In what follows, we briefly review the high-energy regularization schemes which we discuss in the main text. 


\subsection{Lattice Regularization} \label{lattice}

A common way to regularize a quantum field theory is the introduction of a space-time lattice. In this method, the finite lattice spacing $a$ introduces a  momentum cut-off $\Lambda_\mathrm{lattice} \! \propto \! a^{-1}$. 
Lattice regularization  explicitly breaks the parity symmetry of  classical QED$_{2+1}$. This is not a property of the lattice itself, but happens due the mandatory introduction of additional terms in the Lagrangian which prevent artificial gap closings at the high symmetry points of the lattice Brillouin zone. In particular, the Euclidean lattice action $\mathcal{S}_\mathrm{latt}(\bar{\psi},\psi,A)$  is given by \cite{Susskind74,Coste89,capitani2002}
\begin{align} \label{latticeaction}
S_\mathrm{latt} (\bar{\psi},\psi,A)& =-a^3 \sum_{x} \bar{\psi}(x)(D-m)\psi(x) \ ,
\end{align}
where $x=(a n_1, a n_2, an_3)$ with $n_\mu \in \mathbb{Z}$. Moreover, 
\begin{align}
D &= \frac{1}{2} \gamma_\mu^\mathrm{E} (\nabla^\star_\mu + \nabla_\mu) + \frac{1}{2} s a \nabla^\star_\mu \nabla_\mu \ 
\end{align}
is the massless 3 dimensional lattice Dirac operator, $\nabla_\mu$ is the lattice covariant derivative, and  $\gamma_\mu^\mathrm{E}\!=\! \sigma_\mu$ are the Euclidean Dirac matrices. In comparison to continuum QED$_{2+1}$, $S_\mathrm{latt}(\bar{\psi},\psi,A)$ includes the so-called Wilson term, proportional to the Wilson parameter $s=\pm1$. This term acts like a momentum dependent fermion mass and therefore explicitly breaks the parity symmetry of the system. As a consequence, it induces a Chern-Simons term in the effective action, proportional to $\mathrm{sgn}(s)$. Hence, this term acts very similar to the Newtonian $B \vert \mathbf{k} \vert^2$ term in Eq.~\eqref{hamchern}. However, there are two significant differences. One the one hand, the Wilson mass is Lorentz covariant, while the Newtonian mass breaks this symmetry. On the other hand, the Wilson mass vanishes as $a \rightarrow 0$, which is not the case for the  Newtonian mass term as it is a real parameter of the system.


\subsection{Higher Derivative Regularization} \label{higher}

The Lagrangian associated to the higher derivative regularization of QED$_{2+1}$ is given by \cite{Fujikawa85,Nagahama1986b}:
\begin{align}
\mathcal{L}_\mathrm{HD}=\bar{\psi} \left( \im \gamma^\mu \partial_\mu -m \right) \left(1+ \dfrac{\partial^2}{M^2} \right)\psi+e \, \bar{\psi} \gamma^\mu \mathcal{A}_\mu \psi \, ,
\end{align}
where $M$ is a parameter, allowing to remove the higher derivative correction during the renormalization process, $M \rightarrow \infty$. Notice, that by construction, the higher derivative term breaks local gauge invariance. However, this property  is fixed during renormalization \cite{Nagahama1986}. If the higher derivative correction would come with covariant derivatives, it would not reduce the superficial degree of divergence. Hence, the  higher derivative regulator differs significantly form the Newtonian mass in Eq.~\eqref{hamchern}. While the  Newtonian mass term is parity-odd, breaks Lorentz symmetry and renormalizes the Dirac mass in a gauge invariant fashion, the higher derivative term is Lorenz invariant, breaks local gauge symmetry and multiplicates the full non-interacting QED$_{2+1}$ part. As such, it contains a parity even as well as a parity odd contribution. Moreover, it vanishes during renormalization as $M \rightarrow \infty$.  


\subsection{Pauli-Villars Regularization}

In contrast to the two schemes above, the Pauli-Villars regularization adds additional bosonic \mbox{particles} $\chi$ to the classic QED$_{2+1}$ Lagrangian:
\begin{align}
\mathcal{L}_\mathrm{PV}=\bar{\psi} \left( \im \gamma^\mu D_\mu -m \right) \psi +\bar{\chi} \left( \im \gamma^\mu D_\mu -M \right) \chi \ .
\end{align}
Their mass term $M$ breaks parity and therefore induces a Chern-Simons term in the effective action, proportional to sgn($M$). During renormalization the Pauli-Villars field decouples from the theory, $M \rightarrow \infty$. However, it still leaves its trace in the Chern number \cite{Redlich83,Redlich84}. Since this regularization scheme includes an additional particle to ensure an integer quantized DC conductivity, it significantly differs from the Newtonian mass term in Eq.~\eqref{hamchern}.

\newpage

\subsection{Zeta-Function Regularization}

The $\zeta$-function regularization  completely differs from the schemes introduced above\cite{schaposnik96}. It regularizes the generating functional via a certain calculation scheme for the fermion determinant:
\begin{align}
Z[\mathcal{A}]_\mathrm{reg}=\mathrm{det} D[\mathcal{A}] \Big \vert_\mathrm{reg}=\mathrm{e}^{-\frac{d}{ds} \zeta(D[\mathcal{A}],s)} \Big \vert_{s=0} \ .
\end{align}
Here, the Euclidean Dirac operator  with $\gamma_\mu^\mathrm{E}=\sigma_\mu$ and the $\zeta$-function are defined via 
\begin{align}
& D[\mathcal{A}] =\gamma_\mu^\mathrm{E} (\im \partial^\mu+e \mathcal{A}^\mu) + \im m \ ,
\\
& \zeta(D[\mathcal{A}],s) =\mathrm{Tr}\left( D^{-s}[\mathcal{A}] \right) \ . \nonumber
\end{align}
By construction, this scheme is  gauge invariant, but breaks parity symmetry, related to peculiarities during the associated contour integration  (explicit path). It therefore also leads to an integer quantized DC conductivity where one part comes from sgn($m$) and an additional contribution ($\pm 1$) stems from the choice of the integration contour\cite{schaposnik96}.


\section{AC Hall Conductivity}
\label{fullsol}

The general solution of $\im \Pi_\mathrm{CS}^{\mu \nu}(p_0,\mathbf{p}=0)$ in Eq.~\eqref{csint}, is given by the AC Hall conductivity 
\begin{align}
 \sigma_\mathrm{xy}(p_0)=-\dfrac{e^2 }{2 h} & \sum \limits_{s=\pm1} \bigg [ \dfrac{\left(1\!-\!4mB\!-\! \Delta_{p_0} \right) (\mathrm{Ln}_1+\mathrm{Ln}_2)}{4B \Delta_{p_0} \vert p_0 \vert}  \nonumber \\
& \ \,  + \dfrac{\left(1\!-\!4mB\!+\! \Delta_{p_0} \right) (\mathrm{Ln}_3+\mathrm{Ln}_4)}{4B \Delta_{p_0} \vert p_0 \vert} \bigg ] 
\end{align}
where we defined $\Delta_{p_0}=s\sqrt{1-4mB+B^2p_0^2}$ and abbreviate the four logarithms
\begin{align}
\mathrm{Ln}_1 & =  \mathrm{Ln}\left[ -2sB^2 \right] \nonumber \\ \nonumber
\mathrm{Ln}_2 & =   \mathrm{Ln}\left[ s \!   \left( 1 \! - \! 
 4 m B  \! - \!  2 B^2  \vert m \vert \vert p_0 \vert \right)\! \Delta_{p_0}   - s  (1\!-\! 2 m B) \Delta_{p_0}^2  \right]\nonumber \\
 \mathrm{Ln}_3 &=  \mathrm{Ln}\left[ s \left(1\!-\!2mB\!+\!\Delta_{p_0} \right) \right] \nonumber \\ 
\mathrm{Ln}_4 &=  \mathrm{Ln}\left[ 2 s B^2 \!   \left(  \Delta_{p_0}^2 - \Delta_{p_0} \vert B \vert \vert p_0 \vert \right)   \right]  \ .  
\end{align}
Notice, that exactly at $p_0=	 \Delta$, $\Delta_{p_0}$ evaluates to zero. Hence, this quantity encodes the physics stemming from the mass gap apart from the $\Gamma-$point.

\newpage

\bibliographystyle{apsrev4-1}
\bibliography{newBib}

\end{document}